Gamma-Ray Bursts as a Threat to Life on Earth

B.C. Thomas
Department of Physics and Astronomy
Washburn University
1700 SW College Ave.
Topeka, KS 66621
785-670-2144
brian.thomas@washburn.edu

Abstract:
Gamma-ray bursts (GRBs) are likely to have made a number of significant impacts on the Earth during the last billion years. The gamma radiation from a burst within a few kiloparsecs would quickly deplete much of the Earth's protective ozone layer, allowing an increase in solar ultraviolet radiation reaching the surface. This radiation is harmful to life, damaging DNA and causing sunburn. In addition, $NO_2$ produced in the atmosphere would cause a decrease in visible sunlight reaching the surface and could cause global cooling. Nitric acid rain could stress portions of the biosphere, but the increased nitrate deposition could be helpful to land plants. We have used a two-dimensional atmospheric model to investigate the effects on the Earth's atmosphere of GRBs delivering a range of fluences, at various latitudes, at the equinoxes and solstices, and at different times of day. We have estimated DNA damage levels caused by increased solar UVB radiation, reduction in solar visible light due to $NO_2$ opacity, and deposition of nitrates through rainout of $HNO_3$. In this paper I give a concise review of this work and discuss current and future work on extending and improving our estimates of the terrestrial impact of a GRB.

Introduction

Gamma-ray bursts (GRBs) have been recognized as the most powerful explosions in the universe (see e.g. Meszaros 2001; Piran 2005; Bloom *et al.* 2009). There appear to be two classes of GRB, based on duration of the event, which is also correlated with spectral hardness – short duration bursts tend to have harder spectra but less overall energy (Kouveliotou *et al.* 1993; Zhang and Choi 2008). There may also be different populations within these classes, based on luminosity (Chapman *et al.* 2007; Virgili *et al.* 2009; Berger *et al.* 2007; Cenko *et al.* 2008; Chapman et al. 2008). While still under study, long bursts are associated with core-collapse supernovae (Meszaros 2001; Piran 2005; Campana *et al.* 2008), while short bursts may be the result of mergers between compact objects such as neutron stars and black holes (O'Brien & Willingale 2007; Levan 2008).

Starting in the 1990's it was recognized that GRBs, like supernovae and other astrophysical sources of ionizing radiation, could pose a threat to life on Earth over long time scales (Thorsett 1995; Scalo & Wheeler 2002; Melott *et al.* 2004). The radiation from a GRB is highly beamed, and therefore the Earth must fall within that beam in order to be impacted. It has been estimated that the nearest likely long duration burst pointed at the Earth in the last billion years would be on the order of 1 kpc distant (Melott 2004; Dermer & Holmes 2005; Thomas & Melott 2006).

The author and collaborators (see http://kusmos.phsx.ku.edu/~melott/Astrobiology.htm) have performed extensive simulations of the likely impact on the Earth's atmosphere and biosphere by a "typical" long duration GRB. Here I will briefly review this work, describe some recent efforts and discuss some outstanding issues in more accurately quantifying the impact of GRBs on life on Earth. Full details of past work can be found in Thomas *et al.* (2005a,b), Melott *et al.* (2005), Thomas & Melott (2006), Thomas & Honeyman (2008) and Melott & Thomas (2009). In addition, we have explored a wider range of event duration and spectral parameters in order to draw more general conclusions about the impact of astrophysical sources of ionizing radiation (Ejzak *et al.* 2007).

Modeling Parameters

Our simulations have been performed using the Goddard Space Flight Center 2-dimensional atmospheric chemistry and dynamics model, developed by Charles Jackman and others. This code has been used extensively to study ozone changes due to a variety of effects, including supernovae (Geherls *et al.* 2003). The model has 18 latitude bands and 58 altitude bands in log pressure. Further details of the model may be found in Thomas *et al.* (2005b).

Effects discussed below will primarily focus on results obtained for a "standard" long GRB with power $5 \times 10^{44}$ W, with duration 10 seconds, at a distance of 2 kpc, delivering a fluence of 100 kJ m$^{-2}$. We have used the Band spectrum (Band *et al.* 1993), with $E_0 = 187.5$ keV, to compute the atmospheric ionization, which is then used as source of nitrogen oxides in the atmospheric model. In Ejzak *et al.* (2007) we have investigated a broader range of event duration and spectra. These results will be discussed in a general way below.

Terrestrial Effects of a GRB

A GRB within a few parsecs that is directed at the Earth will impact one hemisphere of the planet with a short, but intense blast of high-energy photons. Gamma-rays and X-rays are highly attenuated by the Earth's atmosphere. Therefore, the ground-level effects are primarily indirect. A small fraction of the incident energy reaches the ground as dangerous ultraviolet (UV) radiation (Smith *et al.* 2004), but this is limited in time to the duration of the event, which is at most 10's of seconds for a long burst, and is less than a seconds for a short burst. While it is possible that this flash would affect some organisms, it seems unlikely that a biological catastrophe would result from this effect alone. Of course, for planets with thinner atmospheres the energy deposited at the ground would be greater and more serious effects may be expected (Smith *et al.* 2004; Galante & Horvath 2007). We are concerned here with effects on life on Earth and so will concentrate on the longer-term impacts.

There are three potentially harmful long-term effects of a GRB that follow from changes in atmospheric chemistry (Reid & McAfee 1978). High-energy photons cause dissociation, ionization and ionizing dissociations of $N_2$ and $O_2$ in the atmosphere. Subsequent reactions lead to the formation of nitrogen oxides, most importantly NO and $NO_2$. These compounds catalytically deplete ozone ($O_3$) in the stratosphere, leading to increases in surface-level solar UV over long time periods (years). Secondly, $NO_2$ itself is a brown gas that absorbs strongly in the visible. This may potentially have a climatic effect by reducing solar insolation at the ground, thereby leading to cooling. Third, the atmosphere returns to normal via the removal of nitrogen oxides by way of precipitation of nitric acid ($HNO_3$).

While nitric acid rain can have a negative effect on a variety of organisms (including amphibians), it appears that the amount deposited following a GRB would be small enough so as to not have a serious impact (Thomas & Honeyman 2008). There is a possibility that the deposited nitrate may actually benefit some organisms, particularly land plants that may be nitrogen starved.

It is currently unclear whether cooling due to $NO_2$ opacity is likely to be important. Our results indicate a global average reduction of about 1% in solar visible light fluence, lasting a few years (Melott *et al*. 2005). Larger reductions would occur at the poles, where compounds produced by the burst tend to accumulate (due to poleward transport in the stratosphere). It is possible that this reduction in sunlight could initiate a global cooling event, particularly if the climate is near a "tipping point." To date no simulations of the climate effects have been performed, but our data is available for such work.

The primary effect, then, is the enhanced solar UV at the surface that would follow destruction of stratospheric ozone following a GRB. Globally averaged ozone column density reductions of 25-35% are possible, depending on where the burst occurs in latitude and at what time of year. Reductions of up to 75% at a given location and time are seen for some event cases. The atmosphere recovers on the time scale of years. For our standard fluence, full recovery takes just over a decade. Our results may be compared to values associated with recent anthropogenic reductions in $O_3$. Globally, reductions of 3-5% have been seen, with short lived (weeks) reductions of up to 70% under the so-called ozone hole.

As reported in Thomas *et al.* (2005b) and Ejzak *et al.* (2007), $O_3$ depletion scales roughly as the cube root of the fluence of the event. Additionally, the overall depletion is much more sensitive to the hardness of the spectrum and the total energy deposited than to event duration. For a given fluence and spectrum, the maximum globally averaged ozone depletion is roughly constant for a wide range of duration, from fractions of a second to years, though when that maximum occurs varies.

Biological Effects Following $O_3$ Depletion

A variety of effects on organisms may be expected due to the enhanced surface level solar UV following severe depletion of ozone. The UVB (290-315 nm) part of the spectrum is most affected by $O_3$ levels and is known to damage DNA molecules directly but can also have effects on other biomolecules as well (Vincent and Neale 2000). UVB effects appear at the organism level in ways ranging from lowered metabolic rates and photosynthetic capacity, low growth rate, delayed or arrested cell division and/or death. In multicellular organisms, the effects may cause developmental delay and abnormalities, altered tissue composition and cancer. Biological weighting functions (BWFs) specify wavelength dependence of effectiveness (Coohill 1991; Neale 2000). BWFs are used to weight spectral irradiance to obtain total effective irradiance.

In our work to date, we have used the Setlow DNA damage weighting function (Setlow 1974; Smith 1980) to quantify the biological impact of enhanced UVB irradiance following a GRB. We compute the surface level UVB using the solar spectrum, $O_3$ column densities from the modeling results and the Beer-Lambert relation, which assumes that absorption is more important than scattering.

Using this approach, we find DNA damage of up to 16 times the annual global average of pre-burst values. This level of damage lasts a few months in some areas, particularly at mid-latitudes during the summer. Those locations may experience 5-7 times the normal damage level for a period of several years.

Current and Future Work

It has been proposed that a GRB may have initiated the late Ordovician mass extinction (Melott *et al.* 2004). This proposal was based largely on the water-depth dependence of this extinction, which fits well with a radiation event, since UV is attenuated rather strongly in water and hence would affect organisms which dwell at the top of the water column more strongly than those that live toward the bottom. We have recently completed a study investigating the latitude dependence of our predicted DNA damage values compared to observed latitude dependence of extinction from this period (Melott & Thomas 2009). We find that only a south polar burst can fit the observed trend in extinction with latitude. This allows us to predict that (yet to be measured) extinction rates at northern latitudes should be much smaller, since the effect of a polar burst is almost completely contained within that hemisphere.

Obtaining better estimates of the biological damage following a GRB or other ionizing radiation event is an area of on-going effort. While the Setlow function is fairly generic and a good representation of damage to bare DNA, there are many factors that make the actual effect on organisms and ecosystems much more complicated. Organisms may have shielding, repair mechanisms, avoidance strategies, etc. In addition, extrapolation from DNA damage to ecosystem level effects can be problematic due to the complex inter-relationships involved. We are currently working to improve estimates of damage at the organism level and then to connect that to larger scale effects. These efforts include: performing more accurate calculations of solar UV irradiance in the ocean water column under severely depleted ozone; applying a wider variety of weighting functions; and attempting to model broader ecological effects. We are currently focusing our efforts primarily on phytoplankton in the ocean basins. The ocean accounts for about half of global primary productivity, and, accordingly, half of $CO_2$ fixation and $O_2$ evolution (Behrenfeld et al. 2005). Thus, even modest effects on oceanic productivity could have major repercussions throughout the whole marine food chain and, through atmospheric effects, the global climate.

Along with other sources of ionizing radiation (such as supernovae and soft gamma repeaters), gamma-ray bursts present a real threat to life on Earth and other terrestrial planets in the long term. Such events have likely played a role in the development of life on Earth, and a better understanding of their impacts and rates will be an important component of improving our view of the past, and future, of life on Earth. In a broader context, this understanding can also help to constrain concepts such as the Galactic Habitable Zone and potentially guide efforts such as the Search for Extraterrestrial Intelligence (SETI). More detailed effort is now needed to improve estimates of the rates of GRBs and other events, and to more precisely determine the ecosystem-level effects from a given event.

Acknowledgements

The author's travel costs to attend the ESLAB symposium were funded by an International Travel Grant from Washburn University, as well as the College of Arts and Sciences travel fund. Much of the work reported here was supported by NASA Astrobiology grant NNG04GM41G.